\documentclass{aa}  

\usepackage{graphicx}
\usepackage{tikz}
\usetikzlibrary{arrows.meta}
\usepackage{txfonts}
\begin{document} 
   \title{Analytic theory for the tangential YORP produced by the asteroid regolith}
   %\subtitle{What is that?}
   \author{O. Golubov
          \and
          V. Lipatova
          }

   \institute{Institute for Astronomy, V. N. Karazin Kharkiv National University, Sumska Street 35, Kharkiv 61022, Ukraine\\
              \email{oleksiy.golubov@karazin.ua}
    %     \and
             }
   \date{Received Month ??, 2022; accepted Month ??, 2022}

% \abstract{}{}{}{}{} 
% 5 {} token are mandatory
  \abstract
   {The tangential YORP effect is a radiation pressure torque produced by asymmetric thermal emission by structures on the asteroid surface. As such structures, previous works considered boulders of different shapes lying on the surface of the asteroid.}
   {We study the tangential YORP produced by the rough interface of the asteroid's regolith.}
   {We create an approximate analytic theory of heat conduction on a slightly non-flat sinusoidal surface. We analyze the published data on the small-scale shape of the asteroid (162173) Ryugu and estimate its tangential YORP due to the surface roughness.}
   {We derive an analytic formula that expresses the TYORP of a sinusoidal surface in terms of its geometric and thermal properties. TYORP is maximal at the thermal parameter of the order of unity and for the shape irregularities of the order of the thermal wavelength. Application of this equation to Ryugu predicts TYORP, which is 5-70 times greater than its normal YORP effect.}
   {The contribution of the small-scale regolith roughness to the YORP effect of the asteroid can be comparable to the normal YORP and the tangential YORP produced by boulders. The same theory can describe roughness on the asteroid boulders, thus adding a new term to the previously considered TYORP created by boulders.}


   \maketitle
%________________________________________________________________

\section{Introduction}
\label{sec:introduction}

   The YORP effect is a torque that emerges in small irregularly shaped asteroids due to the light pressure of scattered and re-emitted radiation. This torque changes the rotation rate and obliquity of asteroids \citep{Vokrouhlicky15}. The normal YORP effect, or NYORP, is created by the recoil forces normal to the overall large-scale surface due to the global shape asymmetry of the asteroid \citep{rubincam00}. The tangential YORP effect, or TYORP, is a recoil force dragging the asteroid's surface in the tangential direction due to uneven heat distribution in the small-scale structures on the surface \citep{golubov12}.
   
   TYORP has already been modeled numerically for boulders of different shapes:
   \cite{golubov12} studied TYORP for one-dimensional high walls, \cite{golubov14} and \cite{sevecek16} conducted parametric investigations of TYORP for spherical boulders, and \cite{sevecek15} simulated a boulder of a complex asymmetric shape.
   
   Therefore, up to now only TYORP produced by boulders has been simulated. Still, the surface of asteroids can contain large spaces between boulders covered with fine regolith \citep{murdoch15}, and it is important to estimate this regolith's contribution to TYORP. On the one hand, regolith in general is flatter than boulders, which decreases its TYORP. On the other hand, it can have a thermal parameter closer to unity, which can increase TYORP. Given this, and also a generally bigger area covered by regolith than by boulders, it is uncertain whether boulders or regolith contribute to TYORP more.
   
   A very rough estimate of TYORP produced by regolith has been conducted by \cite{golubov17}. An analytic formula based on multiple simplifying assumptions was used, and it was preliminarily estimated that the regolith could create TYORP of approximately the same order of magnitude as the boulders.
   
   This article aims at a more precise evaluation of TYORP produced by the regolith. It goes beyond the order-of-magnitude estimate of \cite{golubov17} and towards a more realistic analytic theory of TYORP for a surface having the shape of a sinusoidal wave. We prefer a sinusoidal wave over other geometries considered in literature -- concave spherical segments, random Gaussians, fractals
   \citep{davidsson15,rozitis11} -- first, because a sinusoidal wave allows a simpler analytic treatment in terms of the perturbation theory, and second, because the Fourier series makes it very convenient for further generalizations and allows an approximate description of other geometries in terms of the sinusoidal Fourier harmonics.
   
   We present a theory for a sinusoidal wave in Section \ref{sec:analytic} and derive an approximate analytic expression for TYORP.
   In Section \ref{sec:ryugu}, we study the small-scale shape of the surface of (162173) Ryugu.
   It allows us to estimate its TYORP in Section \ref{sec:conclusions}.
   Section \ref{sec:discussion} discusses the physical implications of the proposed theory, its limits of applicability and the possible strength of the effect for other asteroids.
   
%__________________________________________________________________
\section{Analytic computation of TYORP for a sinusoidal surface}
\label{sec:analytic}
    \subsection{Non-dimensionalization of the heat conduction equation}
    
    In the dimensional variables, the heat conduction equation has the following form:
        \begin{equation}
        C\rho\frac{\partial T}{\partial t}=\kappa\left(\frac{\partial^2 T}{\partial X^2}+\frac{\partial^2 T}{\partial Y^2}\right).
        \label{an_pde}
        \end{equation}
    In this equation, $T$ is the temperature, $t$ is the time, $C$ is the heat capacity of the subsurface material, $\rho$ is the density of this material and $\kappa$ is its thermal conductivity. $X$ and $Y$ are the coordinates, defined in such a way that the $X$ axis is pointing eastward along the asteroid surface and the $Y$ axis is directed into the depth of the asteroid, with $Y$ being equal to zero on the surface. The curvature of the surface is assumed to be small, thus it is neglected in the notation of the Laplace operator.
    
    The boundary conditions of the heat conduction equation on the asteroid surface has the following form:
        \begin{equation}
        \left.\kappa\frac{\partial T}{\partial Y}\right|_{Y=0}=\left.\epsilon\sigma T^4\right|_{Y=0}-\alpha \Phi(1-A).
        \label{an_boundary}
        \end{equation}    
    Here, $\epsilon$ is the emissivity of the asteroid, $\sigma$ is the Stefan–Boltzmann constant, $\Phi$ is the solar constant at the asteroid’s orbit, and $A$ is the asteroid's albedo. The dimensionless insolation $\alpha$ is defined as the flux of the solar radiation on the surface normalized by the solar constant.
    The first term in the equation is responsible for the heat flow directed from the depth of the asteroid towards its surface. The next term represents the radiated heat corresponding to the Stefan-Boltzmann law. The last term accounts for the absorption of the solar radiation by the surface of the asteroid.
    
    To simplify the calculations, it is convenient to bring these two equations into dimensionless form. The length scale is defined by the thermal wavelength:
        \begin{equation}
        L_\mathrm{wave}=\sqrt{\frac{\kappa}{C\rho\omega}},
        \label{L_cond}
        \end{equation}
    where $\omega$ is the angular velocity of the asteroid. Characteristic temperature is the temperature of the subsolar point,
        \begin{equation}
        T_0=\sqrt[4]{\frac{(1-A)\Phi}{\epsilon\sigma}}.
        \label{t_0}
        \end{equation}
    After non-dimensionalization, all the essential material properties get incorporated into the thermal parameter, which is defined as:
        \begin{equation} 
        \theta=\frac{\left(C\rho\kappa\omega\right)^{1/2}}{\left((1-A)\Phi\right)^{3/4}\left(\epsilon \sigma \right)^{1/4}}.
        \label{theta}
        \end{equation}
    We express distances in the units of $L_\mathrm{wave}$, temperatures in the units of $T_0$, and time in the units of the asteroid rotation phase. Therefore, we get such dimensionless variables: $\phi=\omega t$, $\tau=T/T_0$, $x=X/L_\mathrm{wave}$ and $y=Y/L_\mathrm{wave}$.
    
    In these notations, the heat conduction equation acquires the following form:
        \begin{equation}
        \frac{\partial\tau}{\partial\phi}=\frac{\partial^2\tau}{\partial x^2}+\frac{\partial^2\tau}{\partial y^2}
        \label{an_pde_dimentionless}
        \end{equation}
    The boundary condition for this equation is: 
        \begin{equation}
        \left.\theta\frac{\partial\tau}{\partial y}\right|_{y=0}=\left.\tau^4\right|_{y=0}-\alpha
        \label{an_boundary_dimentionless}
        \end{equation}
    
    Before solving this equation, let us determine the insolation $\alpha$ that must be substituted into it.
    
    \subsection{Determination of the insolation}
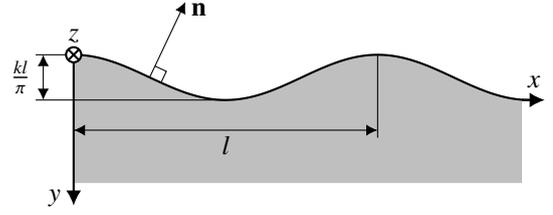
\begin{figure}
        \centering
        \begin{tikzpicture}
        \fill [lightgray, domain=0:5.9, variable=\x] (0,-1.7) -- plot ({\x}, {-0.3+0.3*cos(\x*360/4}) -- (5.9,-1.7) -- cycle;
        \draw[domain=0.1:6,samples=100,thick] plot (\x,{-0.3+0.3*cos(\x*360/4});
        
        \draw[-{Latex[length=1.5mm, width=1.5mm]}] (1,-0.3) -- (1.47,0.7);
        \draw[] (1.07,-0.15)--(1.22,-0.22)--(1.15,-0.37);
        
        \draw[] (4,0)--(4,-1.1);
        \draw[{Latex[length=1.5mm, width=1.5mm]}-{Latex[length=1.5mm, width=1.5mm]}] (0,-1)--(4,-1);
        
        \draw[] (0,0)--(-0.5,0);
        \draw[] (2,-0.6)--(-0.5,-0.6);
        \draw[{Latex[length=1.5mm, width=1.5mm]}-{Latex[length=1.5mm, width=1.5mm]}] (-0.4,0)--(-0.4,-0.6);
        
        \draw[-{Latex[length=2mm, width=2mm]}, thick] (0,-0.1) -- (0,-2);
        \draw[-{Latex[length=2mm, width=2mm]}, thick] (5.9,-0.6) -- (6.2,-0.6);
        \fill[white] (0,0) circle [radius=0.1];
        \draw[thick] (0,0) circle [radius=0.1];
        \draw[thick] (-0.071,0.071)--(0.071,-0.071);
        \draw[thick] (-0.071,-0.071)--(0.071,0.071);
        
        \node[] at (0,0.25) {$z$};
        \node[] at (-0.25,-1.9) {$y$};
        \node[] at (6.05,-0.35) {$x$};
        \node[] at (2,-1.2) {$l$};
        \node[] at (-0.7,-0.3) {$\frac{kl}{\pi}$};
        \node[] at (1.65,0.6) {$\mathbf{n}$};
        
        \end{tikzpicture}
        \caption{Side view of the asteroid surface. The surface has a sinusoidal shape, with the wavelength $l$ and the amplitude $\frac{kl}{2\pi}$, so that the maximum slope is $k$. The $x$ axis is directed along the sine wave, and the $y$ coordinate is measured down from the surface of the asteroid. The normal vector is $\mathbf{n}$.}
        \label{fig:xyz}
    \end{figure}

\begin{figure}
        \centering
        \begin{tikzpicture}
        
        \scope[transform canvas={rotate=-20}]
	    \shade[shading=axis,top color=white,bottom color=black!50,shading angle=90] (0,0) rectangle (1,5.7);
	    \shade[shading=axis,top color=black!50,bottom color=white,shading angle=90] (0.99,0) rectangle (2,5.7);
	    \shade[shading=axis,top color=white,bottom color=black!50,shading angle=90] (1.99,0) rectangle (3,5.7);
	    \shade[shading=axis,top color=black!50,bottom color=white,shading angle=90] (2.99,0) rectangle (4,5.7);
	    \shade[shading=axis,top color=white!0,bottom color=black!50,shading angle=90] (3.99,0) rectangle (5,5.7);
        \endscope
        
        \draw[-{Latex[length=2mm, width=2mm]}, thick] (0,0) -- (6,0);
        \draw[-{Latex[length=2mm, width=2mm]}, thick] (0,0) -- (0,6);
        \draw[-{Latex[length=2mm, width=2mm]}, thick] (0,0) -- (5.63,-2.05);
        \draw[-{Latex[length=2mm, width=2mm]}, thick] (0,0) -- (2.05,5.63);
        
        \draw[{Latex[length=2mm, width=2mm]}-{Latex[length=2mm, width=2mm]}, thick] (5,5.5) -- (6,5.5);
        \draw[{Latex[length=2mm, width=2mm]}-{Latex[length=2mm, width=2mm]}, thick] (5.5,5) -- (5.5,6);
        \fill[black] (5.5,5.5) circle [radius=0.1];
        \node[] at (4.8,5.5) {W};
        \node[] at (6.15,5.5) {E};
        \node[] at (5.5,6.2) {N};
        \node[] at (5.5,4.8) {S};
        
        \fill[white] (0,0) circle [radius=0.1];
        \draw[thick] (0,0) circle [radius=0.1];
        \draw[thick] (-0.071,0.071)--(0.071,-0.071);
        \draw[thick] (-0.071,-0.071)--(0.071,0.071);
        
        \node[] at (0.25,1.4) {$\beta$};
        \node[] at (-0.25,5.9) {$z_1$};
        \node[] at (5.9,0.25) {$x_1$};
        \node[] at (1.7,5.5) {$z$};
        \node[] at (5.5,-1.7) {$x$};
        \node[] at (-0.2,-0.3) {$y_1\equiv y$};
        
        \draw[](0,1.1) arc (90:70:1.1);
        
        \end{tikzpicture}
        \caption{Position of the sine wave on the asteroid surface. The $x_1$ and $z_1$ axes are following the cardinal directions. The $x$ and $z$ axes are linked to the sine wave from Figure \ref{fig:xyz}, and rotated with respect to the cardinal directions by the azimuthal angle $\beta$. The $y$ and $y_1$ axes are both directed downward and coincide with each other.}
        \label{fig:x1y1z1}
    \end{figure}
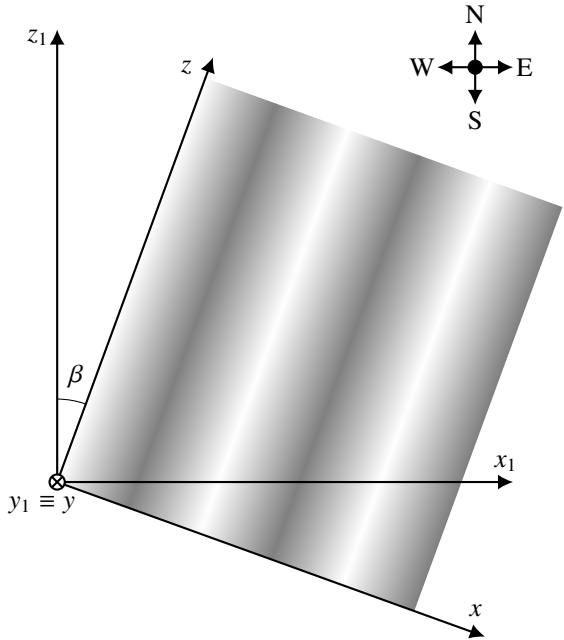

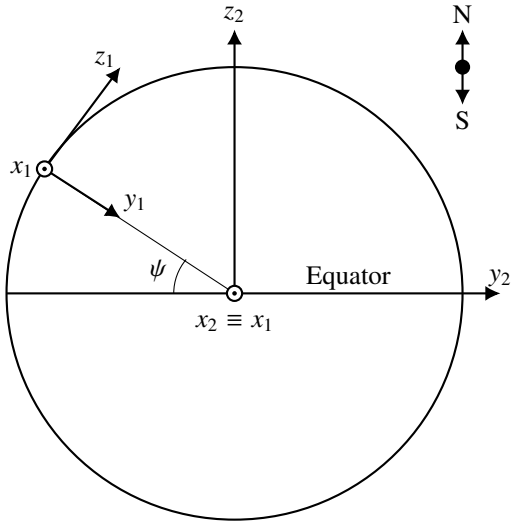
\begin{figure}
        \centering
        \begin{tikzpicture}
        
        \draw[-{Latex[length=2mm, width=2mm]},thick] (2.5,0.5)--(9,0.5);
        \draw[-{Latex[length=2mm, width=2mm]},thick] (5.5,0.6)--(5.5,4);
        \draw[] (5.5,0.5)--(3,2.15);
        \draw[-{Latex[length=2mm, width=2mm]},thick] (3,2.15)--(4,1.5);
        \draw[-{Latex[length=2mm, width=2mm]},thick] (3,2.15)--(4,3.5);
        
        \draw[thick] (5.5,0.5) circle [radius=3];
        
        \fill[white] (5.5,0.5) circle [radius=0.1];
        \draw[thick] (5.5,0.5) circle [radius=0.1];
        \fill[black] (5.5,0.5) circle [radius=0.03];
        
        \fill[white] (3,2.15) circle [radius=0.1];
        \draw[thick] (3,2.15) circle [radius=0.1];
        \fill[black] (3,2.15) circle [radius=0.03];
        
        \node[] at (3.8,3.6) {$z_1$};
        \node[] at (9,0.7) {$y_2$};
        \node[] at (4.2,1.7) {$y_1$};
        \node[] at (5.5,4.2) {$z_2$};
        \node[] at (2.7,2.15) {$x_1$};
        \node[] at (5.5,0.1) {$x_2 \equiv x_1$};
        \node[] at (4.5,0.8) {$\psi$};
        \node[] at (7,0.7) {Equator};
        
        \draw[{Latex[length=2mm, width=2mm]}-{Latex[length=2mm, width=2mm]}, thick] (8.5,3) -- (8.5,4);
        \fill[black] (8.5,3.5) circle [radius=0.1];
        \node[] at (8.5,4.2) {N};
        \node[] at (8.5,2.8) {S};
        
        \draw[](4.7,0.5) arc (180:140:0.7);
        
        \end{tikzpicture}
        \caption{Position of the surface element on the asteroid body. The big circle represents the asteroid, the north pole on the top, the equator in the middle. The axes $x_1$ and $z_1$ are linked to the surface element from Figure \ref{fig:x1y1z1}. The latitude $\psi$ is defined as the angle between the equatorial plane and the global normal to the surface $-y_1$. The axes $x_2$, $y_2$, $z_2$ are defined in the asteroid's body-fixed frame. The axis $z_2$ is aligned with the rotational axis of the asteroid, whereas $x_2$ and $y_2$ are aligned with its equatorial plane and co-rotating with the asteroid.}
        \label{fig:x2y2z2}
    \end{figure}

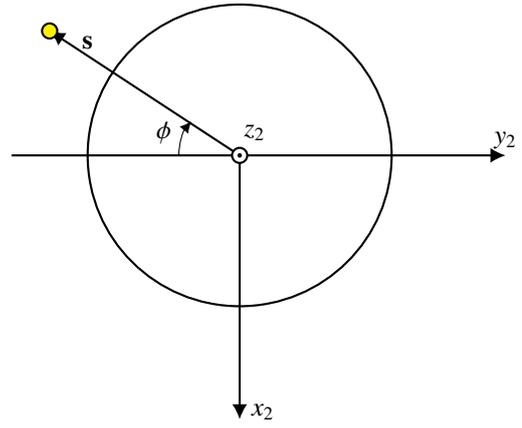
\begin{figure}
        \centering
        \begin{tikzpicture}
        \draw[-{Latex[length=2mm, width=2mm]},thick] (2.5,0.5)--(9,0.5);
        \draw[-{Latex[length=2mm, width=2mm]},thick] (5.5,0.6)--(5.5,-3);
        \draw[-{Latex[length=2mm, width=2mm]},thick] (5.5,0.5)--(3,2.15);
        
        \draw[thick] (5.5,0.5) circle [radius=2];
        
        \fill[white] (5.5,0.5) circle [radius=0.1];
        \draw[thick] (5.5,0.5) circle [radius=0.1];
        \fill[black] (5.5,0.5) circle [radius=0.03];
        
        \fill[yellow] (3,2.15) circle [radius=0.1];
        \draw[thick] (3,2.15) circle [radius=0.1];
        
        \node[] at (9,0.7) {$y_2$};
        \node[] at (3.5,2) {$\mathbf{s}$};
        \node[] at (5.8,-2.9) {$x_2$};
        \node[] at (5.7,0.8) {$z_2$};
        \node[] at (4.5,0.8) {$\phi$};
        
        \draw[-{Latex[length=1.5mm, width=1.5mm]}](4.7,0.5) arc (180:140:0.7);
        
        \end{tikzpicture}
        \caption{Rotation of the Sun in the asteroid's body-fixed frame. The big circle represents the north-pole view of the asteroid from Figure \ref{fig:x2y2z2}. The angle $\phi$ characterizes the time of day on the asteroid. It is measured between the negative $y_2$ axis and the vector $\mathbf{s}$ directed towards the Sun.}
        \label{fig:x2y2z2sun}
    \end{figure}

    The normal to the asteroid surface $\mathbf{n}$ and the vector $\mathbf{s}$ directed towards the Sun are illustrated in Figures \ref{fig:xyz} and \ref{fig:x2y2z2sun}, which are connected by the coordinate transformations shown in Figures \ref{fig:x1y1z1} and \ref{fig:x2y2z2}. In the non-dimensional coordinates, the height of the sinusoidal wave on the surface of regolith (Figure \ref{fig:xyz}) is given by the equation $\frac{kl}{2\pi}\sin(\frac{2\pi x}{l})$, where $k$ is the maximum slope, $l=L/L_\mathrm{cond}$ is the dimensionless wavelength, and $L$ is the dimensional wavelength. When determining the orientation of the normal vector, we assume the surface slopes to be small and thus neglect the terms proportional to $k^2$. The azimuthal angle $\beta$ shows the orientation of the crests of the sine wave with respect to the northward direction on the asteroid surface (Figure \ref{fig:x1y1z1}). The angle $\psi$ is the latitude of the considered surface element on the asteroid (Figure \ref{fig:x2y2z2}). The angle $\phi$ characterizes the position of the Sun with respect to the asteroid (Figure \ref{fig:x2y2z2sun}) and is identical to the dimentionless time in Eqn. (\ref{an_pde_dimentionless}). We assume zero obliquity $\gamma=0$, so that the asteroid's equatorial and orbital planes coincide. In the general case of an arbitrary obliquity, we expect TYORP to be roughly proportional to $1+\cos^2\gamma$ \citep{sevecek16}.
    
    Performing all the coordinate transformations that link Figure \ref{fig:xyz} to Figure \ref{fig:x2y2z2sun}, we obtain the dot product of the surface normal and the unit vector directed towards the Sun,
        \begin{eqnarray}
        \mathbf{ns}=\cos\phi\cos\psi-k\sin\left(\frac{2\pi x}{l}\right)\sin\phi\cos\beta
        \nonumber \\
        + k\sin\left(\frac{2\pi x}{l}\right)\sin\beta\sin\psi\cos\phi.
        \label{an_alpha}
        \end{eqnarray}
    
    The non-dimensional insolation is determined as $\alpha=\mathbf{ns}\,\mathcal{H}(\mathbf{ns})$.
    Here $\mathcal{H}$ is the Heaviside step function, which discards the negative values of $\mathbf{ns}$, since the surface is not illuminated by the Sun at night.
    
    We expand this non-linear expression for $\alpha$ into a Fourier series and retain only its first- and second-order terms: 
        \begin{eqnarray}
        \alpha\approx\frac{\cos\psi}{\pi}+\frac{1}{2}\cos\phi\cos\psi+\frac{k}{\pi}\sin\left(\frac{2\pi x}{l}\right)\sin\psi\sin\beta \nonumber \\ 
        - \frac{k}{2}\sin\left(\frac{2\pi x}{l}\right)\sin\phi\cos\beta+\frac{k}{2}\sin\left(\frac{2\pi x}{l}\right)\sin\beta\sin\psi\cos\phi.
        \label{an_alpha_fourier}
        \end{eqnarray}
    This approximation will decrease the accuracy of our analytic solution, but in most cases it should preserve the correct order of magnitude.
    
    \subsection{Determination of the TYORP force}
    When the heat conduction equation will be solved and the non-dimensional temperature $\tau$ will be found, we will use it to determine the TYORP force. It is worthwhile to write down the expression for TYORP now, to know beforehand which terms of the temperature we will have to retain.
    
    The thermal radiation recoil force acting on the surface area $S$, which is large with respect to the wavelength $L$, but small with respect to the asteroid radius, is expressed as
        \begin{equation}
        F_x=\left\langle-\frac{2}{3c}\epsilon \sigma \left.T^4\right|_{Y=0} n_x \cos\beta\right\rangle_{X,t}S.
        \label{TYORP_force}
        \end{equation}
    The averaging is performed over time $t$ and coordinate $X$. The factor $n_x$ is the $x$ component of the outer normal to the surface, and $c$ is the speed of light. The coefficient 2/3 arises from averaging the recoil force over Lambert's emission indicatrix. The minus sign appears because the recoil pressure acts against the surface normal. 
    
    It is convenient to express the force $F_x$ in terms of the non-dimentional TYORP pressure $p$, defining the latter by the equation
        \begin{equation}
        F_x=\frac{(1-A)\Phi}{c}pS.
        \label{TYORP_force_nondimensionalization}
        \end{equation}
    Then Eqn. (\ref{TYORP_force_nondimensionalization}) gets transformed into the following expression for the non-dimentional TYORP pressure:
        \begin{equation}
        p=-\frac{2}{3}k\cos\beta\left\langle\left.\tau^4\right|_{y=0}\sin{\frac{2\pi x}{l}}\right\rangle_{x,\phi}
        \label{TYORP_pressure_without_dim}
        \end{equation}
    Now the averaging is performed over the rotation phase $\phi$ and the non-dimensional coordinate $x$. The equation $n_x=k \sin(\frac{2 \pi x}{l})$ has been used for the outer normal to the surface (see Figure \ref{fig:xyz}).
        
    \subsection{Series decomposition of the temperature}
    
    Let us now outline the qualitative ideas for our approximate analytic solution of the heat conduction problem, saving all the algebra for the next subsection. Let us write down the insolation from Eqn. (\ref{an_alpha_fourier}) in the following form:
        \begin{equation}
        \alpha=\alpha_0+\varepsilon\sin_x\sin_\phi+\varepsilon\sin_\phi+\varepsilon\sin_x
        \label{decomposition_alpha}
        \end{equation}
    Here $\sin_\phi$ denotes a sinusoidal function of $\phi$ with some unspecified amplitude and phase. Similarly, $\sin_x$ is a sinusoidal function of the horizontal coordinate with an unspecified amplitude, phase, wavelength and azimuthal direction. We will use the factor $\varepsilon$ as a basis to construct a perturbation theory. The temperature will be represented as a series in terms of powers of $\varepsilon$, 
        \begin{equation}
        \tau=\tau_0+\varepsilon\tau_1+\varepsilon^2\tau_2+...
        \label{decomposition_tau}
        \end{equation}
    
    We substitute this temperature into Eqn. (\ref{an_boundary_dimentionless}), and arrive at the boundary condition in the following form:   
        \begin{align}
        \tau_0'+\varepsilon\tau_1'+\varepsilon^2\tau_2'=\alpha_0+\varepsilon\sin_x\sin_\phi+\varepsilon\sin_\phi+\varepsilon\sin_x \nonumber\\
        -(\tau_0^4+4\varepsilon\tau_0^3\tau_1+4\varepsilon^2\tau_0^3\tau_2+6\varepsilon^2\tau_0^2\tau_1^2...).
        \label{decomposition_boundary_full}
        \end{align}
    Here, we assume that primes denote the $y$ derivatives, and that all the temperatures and their derivatives are calculated at the surface of the asteroid.
    By equating the coefficients with the identical powers of $\varepsilon$ in Eqn. (\ref{decomposition_boundary_full}), the boundary condition can be split into a set of equations:     
        \begin{align}
        & \tau_0'=\alpha_0-\tau_0^4, \nonumber\\
        & \tau_1'=\sin_x\sin_\phi+\sin_\phi+\sin_x-4\tau_0^3\tau_1, \nonumber\\
        & \tau_2'=-4\tau_0^3\tau_2-6\tau_0^2\tau_1^2
        \label{decomposition_eps_set}
        \end{align}
    
    The heat conduction equation
    Eqn. (\ref{an_pde_dimentionless}) can be similarly decomposed into powers of $\varepsilon$. As a result, we assume that all the terms $\tau_0$, $\tau_1$, $\tau_2$ individually satisfy Eqn. (\ref{an_pde_dimentionless}).
    
    From the first line of Eqn. (\ref{decomposition_eps_set}), a constant solution for $\tau_0$ can be found. Then from the second line of Eqn. (\ref{decomposition_eps_set}) and the heat conduction equation Eqn. (\ref{an_pde_dimentionless}), a solution for $\tau_1$ is obtained, which appears to repeat the form of the second line of Eqn. (\ref{decomposition_eps_set}), $\tau_1\sim\sin_x\sin_\phi+\sin_\phi+\sin_x$. (The phases and amplitudes of the $\sin$ terms are different from the ones in Eqn. (\ref{decomposition_tau}), but we do not pay attention to them in this qualitative analysis.) Next, $\tau_1$ must be squared to be substituted into the third line of Eqn. (\ref{decomposition_eps_set}), $\tau_1^2\sim\sin_x^2\sin_\phi^2+\sin_x\sin_\phi^2+\sin_\phi^2+\sin_x^2+\sin_x^2\sin_\phi$. Making use of the addition formulas of trigonometry, we can re-write it in a shorthand notation as $\tau_1^2\sim(1+\sin_{2x})(1+\sin_{2\phi})+\sin_x(1+\sin_{2\phi})+(1+\sin_{2\phi})+(1+\sin_{2x})+(1+\sin_{2x})\sin_\phi$, where ``2'' in the lower indexes corresponds to the trigonometric functions of doubled arguments. Finally, opening the brackets and again ignoring the constant multipliers, we get $\tau_1^2\sim 1+\sin_{2x}+\sin_{2\phi}+\sin_{2x}\sin_{2\phi}+\sin_x+\sin_\phi+\sin_x\sin_{2\phi}+\sin_\phi\sin_{2x}$. This $\tau_1^2$ is substituted into the third line of Eqn. (\ref{decomposition_eps_set}), producing a solution of a similar form, $\tau_2\sim 1+\sin_{2x}+\sin_{2\phi}+\sin_{2x}\sin_{2\phi}+\sin_x+\sin_\phi+\sin_x\sin_{2\phi}+\sin_\phi\sin_{2x}$.
    
    All the components of $\tau$ must be substituted into the expression for TYORP pressure $p\sim\langle(\tau_0^4+4\varepsilon\tau_0^3\tau_1+4\varepsilon^2\tau_0^3\tau_2+6\varepsilon^2\tau_0^2\tau_1^2...)\sin_x\rangle_{x,\phi}$. Only the terms containing trigonometric functions of neither $x$ nor $\phi$ survive after averaging. The result is thus produced by the terms $\sin_x$ in $4\varepsilon\tau_0^3\tau_1$, $4\varepsilon^2\tau_0^3\tau_2$, and $6\varepsilon^2\tau_0^2\tau_1^2$. This finishes the problem of finding the highest-order contribution to $p$ in terms of $\varepsilon$.
        
    The catch is that $\varepsilon$ is not small, but is rather of the order of unity. Still, we can expect that the equations that are precise for $\varepsilon\ll 1$ give a reasonable order-of-magnitude estimate for $\varepsilon\sim 1$, although they are not necessarily precise at any region of the parameter space. This approach is the same as the one, which was used by \cite{golubov17} and lead to a good agreement between the theory and simulations.
        
    One more thing to be pointed out from this qualitative analysis is additivity of waves on the regolith surface. Consider several different sinusoidal spacial waves with non-commensurable wavelengths in the right-hand side of Eqn. (\ref{decomposition_alpha}). These terms from Eqn. (\ref{decomposition_alpha}) will move to the subsequent equations and mix with each other when expressions get squared or multiplied. Still, because of the non-commensurability of the wavelengths, the terms independent of the spacial coordinates will arise only from products of terms with the same wavelengths. And these are the only ones which create a non-vanishing mean TYORP pressure. It implies that the TYORP pressure is created by each sinusoidal harmonics individually, so that different harmonics with non-commensurable wavelengths do not interfere, and the total TYORP is the sum of TYORPs of different harmonics. Though the case of commensurable wavelengths requires a more sophisticated consideration.
    
    After having outlined the perturbation theory on the qualitative level, let us devote the next two subsections to the quantitative realization of the suggested plan.
    
    \subsection{Approximate solution for the temperature}
    We solve the heat conduction equation, treating the non-dimensional insolation $\alpha$ in Eqn. (\ref{an_alpha_fourier}) perturbatively.
    In the zeroth approximation, we take only the first term in Eqn. (\ref{an_alpha_fourier}). It is a constant mean insolation, which leads to a constant solution for the non-dimensional temperature, $\tau_0=\sqrt[4]{\frac{\cos\psi}{\pi}}$
    
    In the next approximation, we add the remaining terms in Eqn. (\ref{an_alpha_fourier}), and find the corrected expression for the temperature, $\tau=\tau_0+\tau_1$, which solves the heat conduction equation Eqn. (\ref{an_pde_dimentionless}) with the boundary condition Eqn. (\ref{an_boundary_dimentionless}). As $\tau_0$ by itself also solves the heat conduction equation, the same must be true for $\tau_1$: 
        \begin{equation}
        \frac{\partial\tau_1}{\partial\phi}=\frac{\partial^2\tau_1}{\partial x^2}+\frac{\partial^2\tau_1}{\partial y^2}
        \label{an_pde_1}
        \end{equation}
    In the boundary condition Eqn. (\ref{an_boundary_dimentionless}), we assume $\tau^4\approx \tau_0^4+4\tau_0^3\tau_1$ and neglect the higher-order terms. The principal term of $\alpha$ cancels $\tau_0^4$, and the boundary condition for $\tau_1$ acquires the following form:    
        \begin{eqnarray}
        \left.\theta\frac{\partial\tau_1}{\partial y}\right|_{y=0}-\left.4\tau_0^3\tau_1\right|_{y=0}= \frac{1}{2}\cos\psi\cos\phi \nonumber \\ -\frac{k}{\pi}\sin\left(\frac{2\pi x}{l}\right)\sin\beta\sin\psi \nonumber \\
        +\frac{k}{2}\sin\left(\frac{2\pi x}{l}\right)\cos\beta\sin\phi -\frac{k}{2}\sin\left(\frac{2\pi x}{l}\right)\sin\beta\sin\psi\cos\phi
        \label{an_boundary_1}
        \end{eqnarray}
    The solution of this inhomogenous linear problem can be found as a sum of solutions corresponding to individual inhomogeneities in the right-hand side of Eqn. (\ref{an_boundary_1}). 
    Each inhomogeneity has a sinusoidal dependence on $\phi$ and $x$, which in terms of complex variables can be treated as an exponential dependence. We allow for an exponential dependence on $y$ as well, and substitute such a product of three exponents into Eqn. (\ref{an_pde_1}). It turns the partial differential equation into a quadratic algebraic equation for the unknown factor standing in front of $y$ in the exponent. We choose only such roots of this quadratic equation, which correspond to the exponential decrease of $\tau_1$ at large depth, not exponential increase. Then the pre-exponential factor is chosen to satisfy Eqn. (\ref{an_boundary_1}). Performing this procedure for each inhomogeneity in (\ref{an_boundary_1}), we find the following solution for $\tau_1$:
        \begin{eqnarray}
        \tau_1=e^{-\frac{y}{\sqrt{2}}}\left(a_1\cos\left({\phi-\frac{y}{\sqrt{2}}}\right)+a_2\sin\left({\phi-\frac{y}{\sqrt{2}}}\right) \right) \nonumber \\
        +b_1 \exp\left({-\frac{2\pi y}{l}}\right)\sin\left({\frac{2\pi x}{l}}\right) \nonumber \\
        +e^{-\mu y} \Bigg{(} c_1\cos\phi\cos\left({\frac{2\pi x}{l}-\nu y}  \right) + c_2\cos\phi\sin\left({\frac{2\pi x}{l}-\nu y}  \right) \nonumber \\
        + c_3\sin\phi\cos\left({\frac{2\pi x}{l}-\nu y}  \right) + c_4\sin\phi\sin\left({\frac{2\pi x}{l}-\nu y}\right) \Bigg{)}
        \label{tau_1}
        \end{eqnarray}
    The coefficients entering this solution are expressed as
        \begin{eqnarray}
        a_1&=&a_2\left(1+\frac{4\sqrt{2}\tau^3_0}{\theta}\right)=\frac{\left(1+\frac{4\sqrt{2}\tau^3_0}{\theta}\right)\cos\psi}{\sqrt{2}\theta\left(1+\left(1+\frac{4\sqrt{2}\tau^3_0}{\theta}\right)^2\right)} \nonumber\\
        b_1&=&\frac{k}{\pi\left(4\tau^3_0+\frac{2\pi\theta}{l}\right)}\nonumber\\
        c_1&=&-c_3\frac{\cos\beta}{\sin\beta\sin\psi}=\frac{\frac{-k}{2\theta\nu}\sin\beta\sin\psi}{1+\left(\frac{\mu}{\nu}+\frac{4\tau^3_0}{\nu\theta}\right)^2}\nonumber\\
        c_2&=&-c_4\frac{\cos\beta}{\sin\beta\sin\psi}=\frac{\frac{k}{2\theta\nu}\sin\beta\sin\psi\left(\frac{\mu}{\nu}+\frac{4\tau^3_0}{\nu\theta}\right)}{1+\left(\frac{\mu}{\nu}+\frac{4\tau^3_0}{\nu\theta}\right)^2}
        \label{tau2_coefficients}
        \end{eqnarray}
    The constants $\mu$ and $\nu$ in these equations are defined as
        \begin{eqnarray}
        \mu=\frac{1}{\sqrt{2}}\left(\sqrt{1+\left(\frac{2\pi}{l}\right)^4}+\left(\frac{2\pi}{l}\right)^2\right)^{\frac{1}{2}}\nonumber\\
        \nu=\frac{1}{\sqrt{2}}\left(\sqrt{1+\left(\frac{2\pi}{l}\right)^4}-\left(\frac{2\pi}{l}\right)^2\right)^{\frac{1}{2}}
        \end{eqnarray}
    
    Next, we implement the second correction to the temperature, $\tau=\tau_0+\tau_1+\tau_2$. In a similar way, we obtain the following equation and the boundary condition for $\tau_2$:
        \begin{equation}
        \frac{\partial\tau_2}{\partial\phi}=\frac{\partial^2\tau_2}{\partial x^2}+\frac{\partial^2\tau_2}{\partial y^2}
        \label{an_pde_2}
        \end{equation}
        \begin{equation}
        \left.\theta\frac{\partial\tau_2}{\partial y}\right|_{y=0}-\left.4\tau_0^3\tau_2\right|_{y=0}= \left.6\tau_0^2\tau_1^2\right|_{y=0}
        \label{an_boundary_2}
        \end{equation}
    Solving this equation at full length would be very tedious, and would produce a very cumbersome answer. But we will see in the next subsection, that only one particular term from $\tau_2$ is needed to estimate TYORP.

\subsection{Derivation of the TYORP force from the approximate expression for the temperature}
        \begin{figure*}
        \begin{center}
        \includegraphics[width=.49\textwidth]{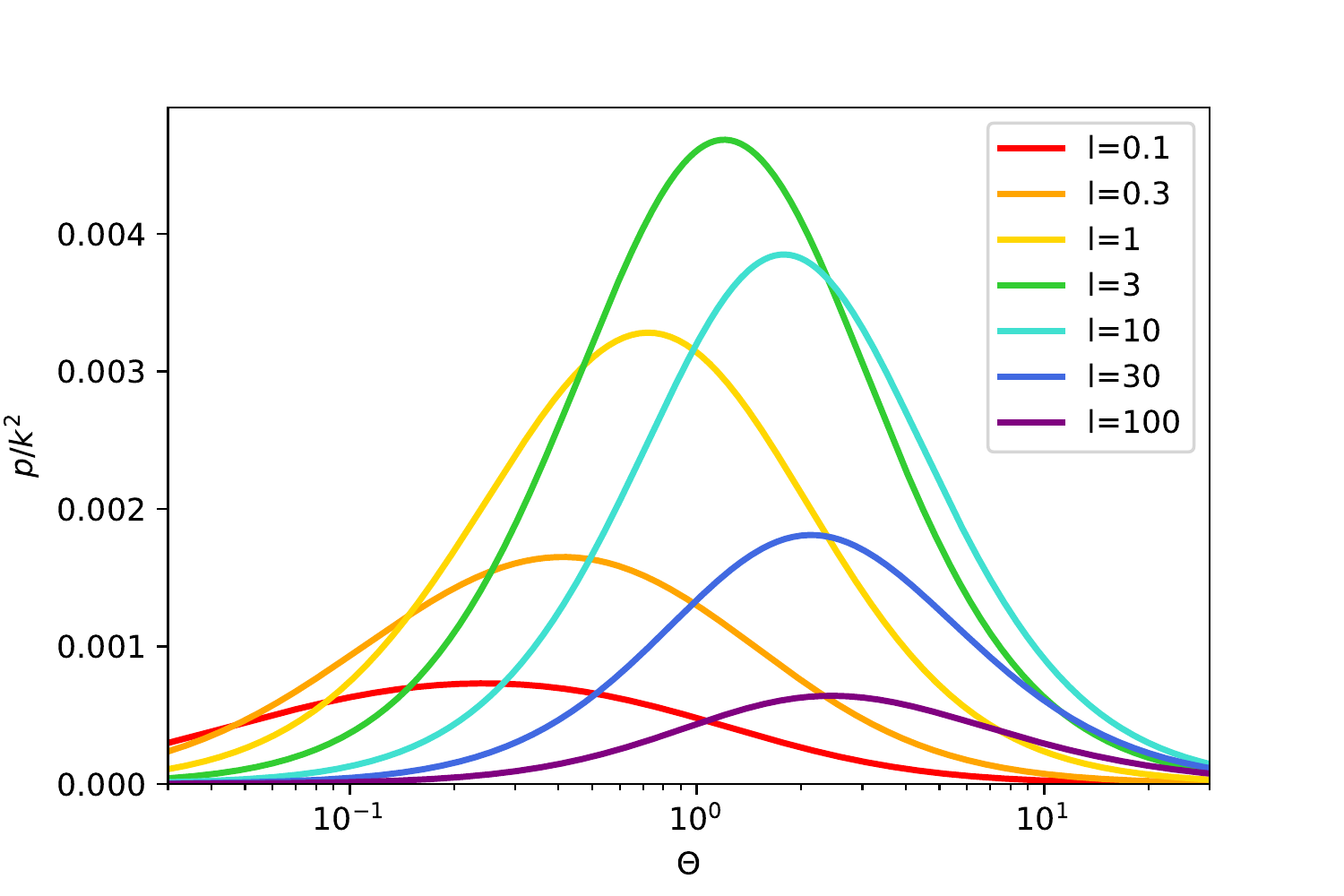}
        \includegraphics[width=.49\textwidth]{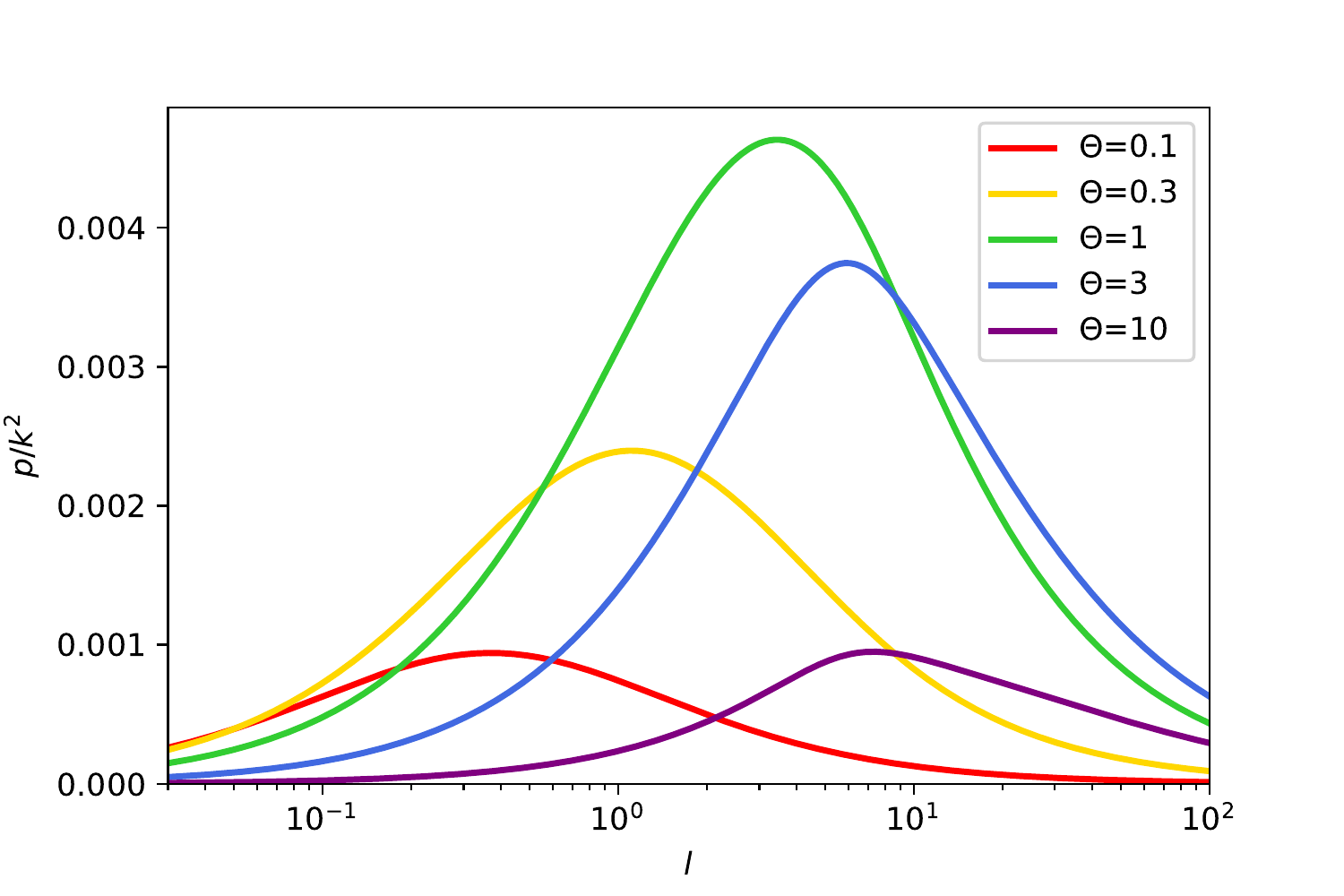}
        \caption{The non-dimensional TYORP pressure determined by Eqn. (\ref{an_pressure_final}) as a function of $l$ and $\theta$. Left panel: Dependence on $\theta$ for several fixed values of $l$. Right panel: Dependence on $l$ for several fixed values of $\theta$. The other parameters are $\psi=0$, $\beta=0$.} 
        \label{fig:tyorp}
        \end{center}
        \end{figure*}
        
        \begin{figure}
        \begin{center}
        \includegraphics[width=.49\textwidth]{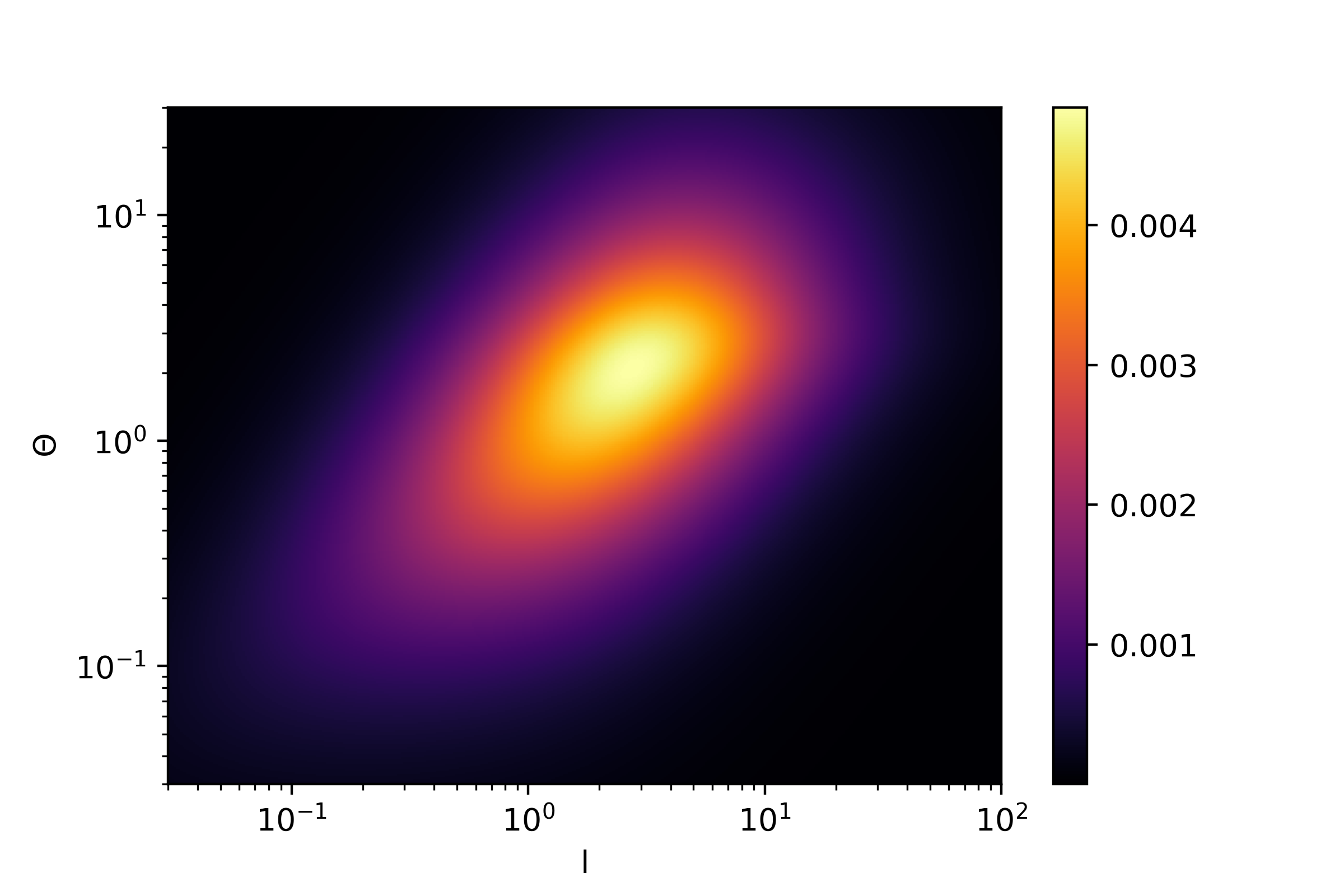}
        \caption{The color map showing the non-dimensional TYORP pressure determined by Eqn. (\ref{an_pressure_final}) as a function of $l$ and $\theta$. The other parameters are $\psi=0$, $\beta=0$.} 
        \label{fig:tyorp2d}
        \end{center}
        \end{figure}
        
        \begin{figure*}
        \begin{center}
        \includegraphics[width=.49\textwidth]{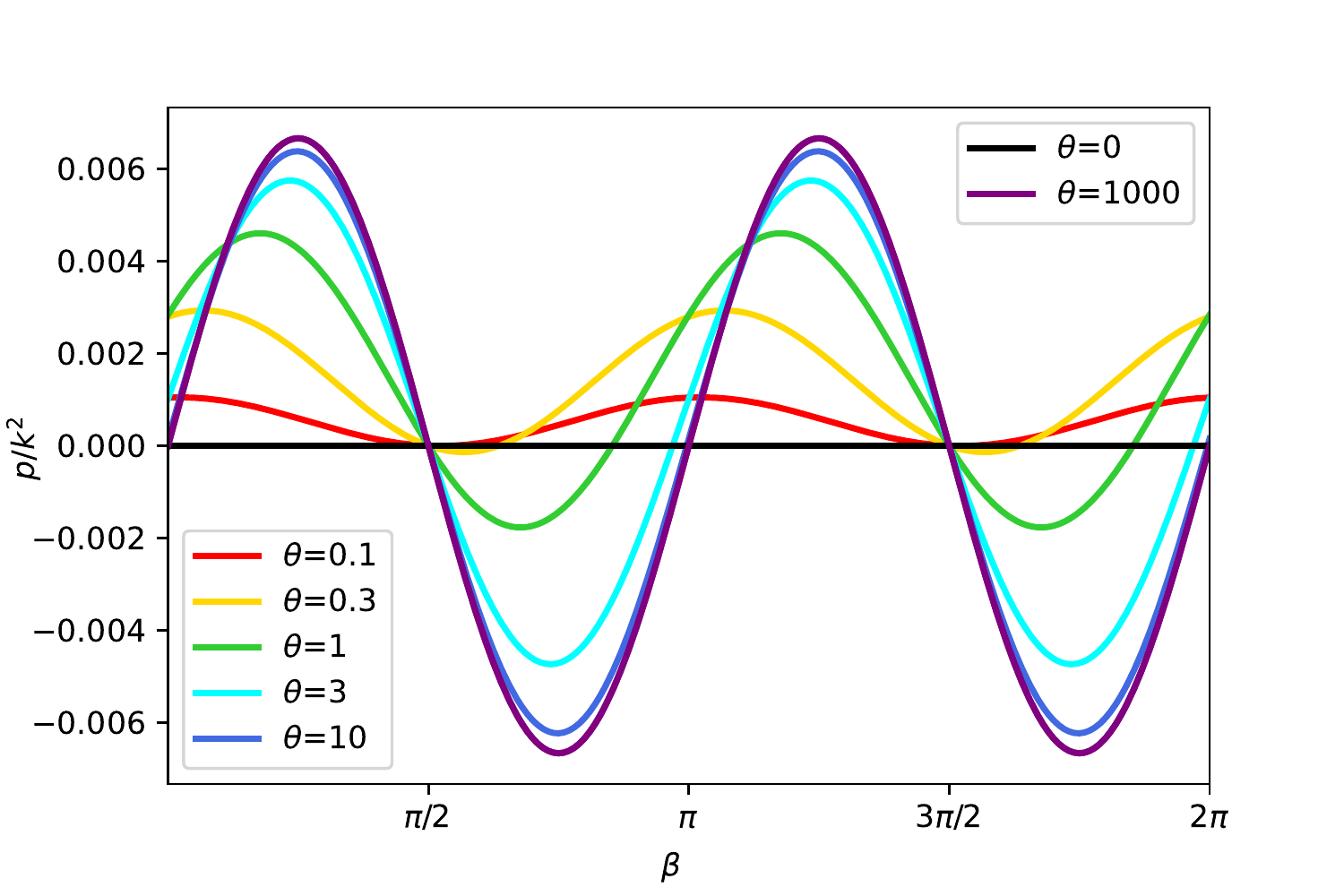}
        \includegraphics[width=.49\textwidth]{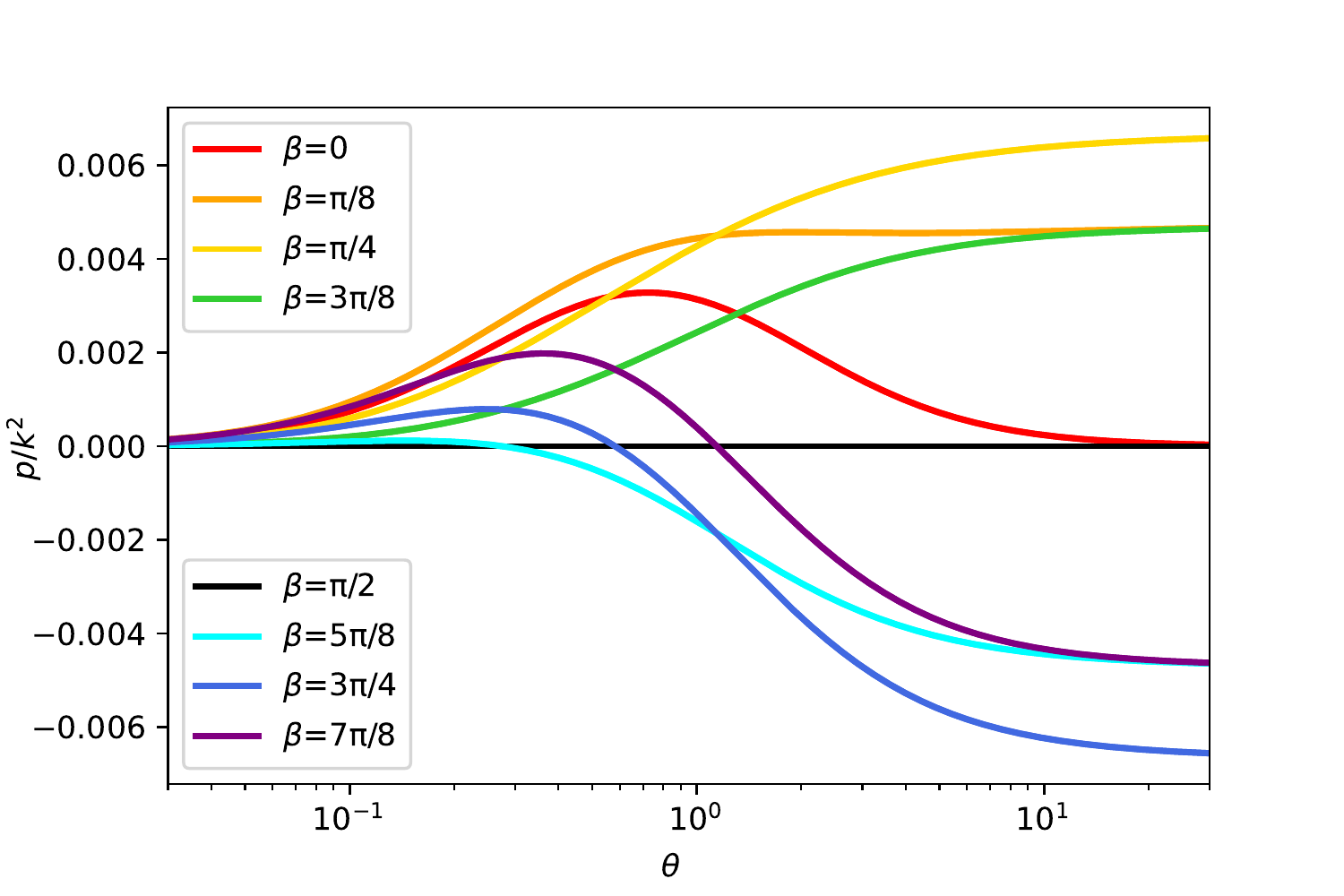}
        \caption{Illustration of the dependence of the non-dimensional TYORP pressure determined by Eqn. (\ref{an_pressure_final}) on the azimuthal angle $\beta$. Left panel: Dependence on $\beta$ for several fixed values of $\theta$. Right panel: Dependence on $\theta$ for several fixed values of $\beta$. The other parameters are $l=1$, $\psi=0$.} 
        \label{fig:tyorp-beta}
        \end{center}
        \end{figure*}
        
    We substitute the approximate expression for the non-dimensional temperature $\tau=\tau_0+\tau_1+\tau_2$ into the expression for the TYORP force given by Eqn. (\ref{TYORP_pressure_without_dim}), neglecting the terms that contain high powers of $\tau_1$ and $\tau_2$:    
        \begin{equation}
        p=-\frac{2}{3}k\cos\beta\left\langle\left(\tau^4_0+4\tau^3_0\tau_1+6\tau^2_0\tau^2_1+4\tau^3_0\tau_2\right)\sin{\frac{2\pi x}{l}}\right\rangle_{x,\phi}
        \label{pressure_via_tau_series}
        \end{equation} 
    The $\tau^4_0$ term gives zero when averaged over $x$. In all other terms of $\tau^4$ only the ones independent of $\phi$ and proportional to $\sin{\frac{2\pi x}{l}}$ survive after averaging.
     
    The only surviving term in the expression for pressure, which comes from $4\tau^3_0\tau_1$, is the one proportional to $b_1$,
        \begin{eqnarray}
        p_\beta=-\frac{k^2 \pi \tau_0^2 \theta^3 \left(4 \tau_0^3 + \mu \theta\right)\left(4 \sqrt{2}\tau_0^3 +  \theta \right) }{2 \sqrt{2} \left(2 l \tau_0^3 + \pi \theta\right) \left(16 \tau_0^6 +
        4 \sqrt{2} \tau_0^3 \theta + \theta^2\right)}\nonumber \\        \times\frac{\cos\psi\sin\psi\cos\beta\sin\beta}{16 \tau_0^6 + 
        8 \mu \tau_0^3 \theta + \left(\mu^2 + \nu^2\right) \theta^2}
        \label{pressure_tau1}
        \end{eqnarray}
    Its physical meaning is that the southern slopes of the sinusoidal mounds are warmer and radiate more heat to the west or east, depending on the azimuth $\beta$.
    Although this term can be the dominating one for small patches of the surface, we do not expect it to produce any significant TYORP for the asteroid as a whole. The reason is the absence of any asymmetry in the azimuth $\beta$ between the north-east and north-west orientations of the waves, leading to a near-zero average of Eqn. (\ref{pressure_tau1}) over the asteroid surface. We will thus disregard this term in our further consideration.
     
    After squaring Eqn. (\ref{tau_1}), one can see that the only component of $6\tau^2_0\tau^2_1$ surviving after averaging in Eqn. (\ref{pressure_via_tau_series}) is:
        \begin{equation}
        p_1=-2k\cos\beta\tau^2_0\left(a_1 c_2+a_2 c_4\right)
        \label{pressure_tau1_squared}
        \end{equation}
        
    From the form of the last term in Eqn. (\ref{pressure_via_tau_series}), $4\tau^3_0\tau_2$, we can conclude, that the only component of $\tau_2$ contributing to TYORP is the one proportional to $\sin{\frac{2\pi x}{l}}$ and independent of $\phi$. From Eqs. (\ref{an_pde_2}) and (\ref{an_boundary_2}) we see that such terms in $\tau_2$ are produced by the corresponding terms in $\tau_1^2$ that are also proportional to $\sin{\frac{2\pi x}{l}}$ and independent of $\phi$. We search for this term in the form of the product of exponents depending on $x$ and $y$. We substitute the product of the exponents into Eqn. (\ref{an_pde_2}), and thus find the dependence of the studied term on $y$. Then we substitute this term into the boundary condition Eqn. (\ref{an_boundary_2}) to find the pre-exponential factor. The resulting expression of the only relevant term in $\tau_2$ is the following:
        \begin{equation}
        \tau_2=-\frac{6\tau^2_0\left(a_1 c_2+a_2 c_4\right)}{\frac{2\pi\theta}{l}+4\tau^3_0}\sin\left(\frac{2\pi x}{l}\right)\exp{\left(-\frac{2\pi y}{l}\right)}
        \label{pressure_tau2}
        \end{equation}
    Finally, we substitute this expression into Eqn. (\ref{pressure_via_tau_series}), and find the contribution of $\tau_2$ to TYORP:
        \begin{equation}
        p_2=8k\cos\beta\tau^3_0\frac{\tau^2_0\left(a_1 c_2+a_2 c_4\right)}{\frac{2\pi\theta}{l}+4\tau^3_0}
        \label{an_pressure_4}
        \end{equation}
        
    Adding up Eqs. (\ref{pressure_tau1_squared}) and (\ref{an_pressure_4}), and substituting the coefficients $a_1$, $a_2$, $c_2$, and $c_4$ from Eqn. (\ref{tau2_coefficients}), we derive the final expression for the non-dimensional TYORP pressure:   
        \begin{eqnarray}
        p=\frac{k^2 \pi \tau_0^2 \theta^2 \left(4 \tau_0^3 + \mu \theta\right)}{2 \sqrt{2} \left(2 l \tau_0^3 + \pi \theta\right) \left(16 \tau_0^6 +
        4 \sqrt{2} \tau_0^3 \theta + \theta^2\right)}\nonumber \\
        \times\frac{\cos\psi\cos^2\beta}{16 \tau_0^6 + 
        8 \mu \tau_0^3 \theta + \left(\mu^2 + \nu^2\right) \theta^2}
        \label{an_pressure_final}
        \end{eqnarray}
    According to this equation, TYORP depends on the following physical parameters: the thermal parameter $\theta$, the non-dimensional wavelength $l$, the azimuth $\beta$, the latitude $\psi$, and maximal slope $k$. The remaining three values $\mu$, $\nu$, and $\tau_0$ are expressed in terms of other parameters.
    
    Dependence of $p$ on $\theta$ and $l$ as given by Eqn. (\ref{an_pressure_final}) is illustrated in Figures \ref{fig:tyorp}--\ref{fig:tyorp-beta}.
    The two panels of Figure \ref{fig:tyorp} show the dependence on $\theta$ and $l$ separately. Figure \ref{fig:tyorp2d} color codes $p$ as a function of $\theta$ and $l$ simultaneously, so that the lines in the left-hand side and the right-hand side of Figure \ref{fig:tyorp2d} correspond to individual vertical and horizontal cross-sections of Figure \ref{fig:tyorp} respectively. We see that TYORP is maximal at $l\sim\theta\sim 1$, and rapidly decays at smaller and bigger $l$ or $\theta$.
    
    The trends shown in Figure \ref{fig:tyorp} are similar to the one presented in Figure 8 of \cite{golubov17}, if $l$ from this article is equated to $\lambda$ from \cite{golubov17}, and $p/k^2$ from this article is equated to $p/\alpha^2$ from \cite{golubov17}. The maximum $p/\alpha^2$ from \cite{golubov17} is about 0.012, which is almost 2.5 times larger than the maximum in Figures \ref{fig:tyorp} and \ref{fig:tyorp2d}.
    
    Figure \ref{fig:tyorp-beta} illustrates the dependence of the total TYORP force $p_\beta+p$ (Eqs. (\ref{pressure_tau1}) and (\ref{an_pressure_final}) combined) on $\beta$ and $\theta$.
    Both $p_\beta$ and $p$ are periodic in $\beta$ with the period $\pi$, but $p$ vanishes at $\beta=\frac{\pi}{2}$ and reaches maximum at $\beta=0$, whereas $p_\beta$ vanishes at $\beta=0$ and $\beta=\frac{\pi}{2}$, reaches maximum at $\beta=\frac{\pi}{4}$ and minimum at $\beta=\frac{3\pi}{4}$. At $\theta\rightarrow 0$, both $p_\beta$ and $p$ vanish. At $\theta\rightarrow\infty$, $p$ vanishes and $p_\beta$ reaches maximum.
    Despite the rich physics and beauty of Figure \ref{fig:tyorp-beta}, it presents mostly an academic interest, because $p_\beta$ vanishes after averaging over $\beta$, and thus only $p$ is important for the global TYORP.
    
\section{Analysis of the shape of (162173) Ryugu}
\label{sec:ryugu}

    As a sample of the asteroid surface, we used the cross-section of a region on the asteroid (162173) Ryugu determined by the MASCOT lander of the \textit{Hayabusa2} space mission \citep{scholten19,otto21}. It has the length 40 cm and the spacial resolution 3 mm.
    
    We want to determine the typical maximum slope $k(l)$ corresponding to each wavelength $l$, in order to substitute it into Eqn. (\ref{an_pressure_final}). For this, we need a measure of curvature of the surface at each $l$. As a simple measure of curvature we adopt the following function:
        \begin{equation}
        \lambda(h)=\left\langle \left(y \left(x+h \right)-2y \left(x+\frac{h}{2} \right)+y(x) \right)^2\right\rangle_x
        \label{lipatogram_def}
        \end{equation}
    This function is proportional to the mean value of the square of the second-order divided difference for the height as a function of the horizontal coordinate $x$, with $h$ being the distance scale at which the divided difference is computed. If the height $y(x)$ has a constant slope, the second-order divided difference vanishes, and $\lambda(h)=0$. If the surface is sinusoidal with the wavelength $l$ and the maximum slope $k$, i.e. $y(x)=\frac{kl}{2\pi}\sin{ \left(\frac{2\pi x}{l} \right)}$, it can be easily shown that
        \begin{equation}
        \lambda(h)=2 \left(\frac{kl}{\pi} \right)^2\sin^4{ \left(\frac{\pi h}{2l} \right)}.
        \label{lipatogram_sin}
        \end{equation}
    The maximum contribution to $\lambda(h)$ is produced by waves with such wavelengths $l$ that $\frac{\pi h}{2l}=\frac{\pi}{2}+\pi n$, where $n\in \mathbb{N}_0$, that is $l=h/(2n+1)$. As the terms of this series decay rapidly with $n$, we approximate the entire series by its first term. Thus we equate the experimentally determined $\lambda(l)$ from Eqn. (\ref{lipatogram_def}) to the theoretical $\lambda(l)$ from Eqn. (\ref{lipatogram_sin}) for the same wavelength $l$. This results into the following expression for the maximal slope of a sinusoidal wave of a given wavelength:
        \begin{equation}
        k(l)=\frac{\pi}{l}\sqrt{\frac{\lambda(l)}{2}}
        \label{aaa}
        \end{equation}
    
\begin{figure}
        \begin{center}
        \includegraphics[width=.49\textwidth]{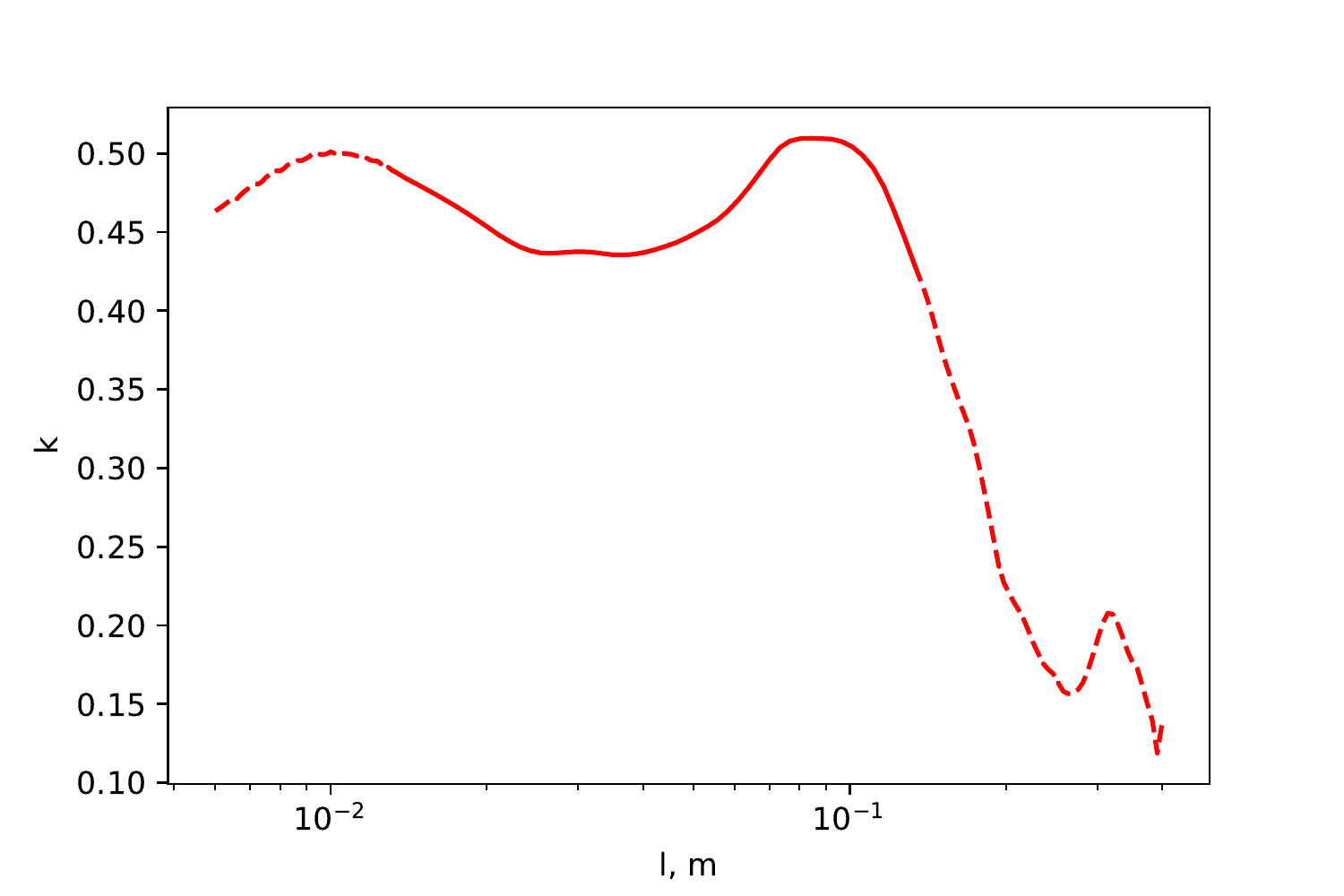}
        \caption{The effective slope of sinusoidal waves of different wavelengths, computed via Eqn. (\ref{aaa}) for the shape of asteroid (162173) Ryugu presented in \cite{otto21}.} 
        \label{fig:k}
        \end{center}
        \end{figure}

    We apply this equation to the cross-section of Ryugu from Figure 2c in \cite{otto21}. The resulting slope $k(l)$ as a function of the wavelength is shown in Figure \ref{fig:k}. The plot shows the entire range from 3 mm (the spacial resolution of the original cross-section) to 40 cm (the total length of the studied cross-section). Close to these limits the measured $k(l)$ has a lower accuracy: due to the loss of small-scale features when $l$ is close to the spatial resolution and due to the insufficient statistics when $l$ is close to the maximal length. We mark these less certain regions with dashed lines in Figure \ref{fig:k}.
    
    This slope $k$ must be substituted into Eqn. (\ref{an_pressure_final}) to get the TYORP pressure for the asteroid. Note, that here we assumed that the waves have zero azimuth, $\beta=0$. For a different azimuth, the factor $\cos\beta$ will appear in the denominator of the right-hand side of Eqn. (\ref{aaa}), but it will be cancelled by the $\cos\beta$ in the numerator of Eqn. (\ref{an_pressure_final}). Another $\cos\beta$ will appear in the argument of the left-hand side of Eqn. (\ref{aaa}), but this horizontal stretching of Figure \ref{fig:k} will leave the near-horizontal part of the plot at the same level and will not effect the resulting TYORP. Thus we just ignore the factors $\cos\beta$ everywhere in our estimate and assume $\beta=0$.
        
\section{Results}
\label{sec:conclusions}

    Finally, the TYORP pressure must be integrated over the surface of the asteroid. We assume the asteroid to be spherical, to have a zero obliquity and a circular orbit. In the dimensional units, the resulting TYORP torque is
        \begin{equation}
        T_\omega=\frac{(1-A)\Phi R^3}{c}p_0 2\pi \int_{-\frac{\pi}{2}}^{\frac{\pi}{2}} \cos^3{\psi} d\psi
        \label{torque_dimentional}
        \end{equation}
    The lower index $\omega$ marks the component of the torque that changes the rotation rate $\omega$ of the asteroid. In this equation, $R$ is the radius of the asteroid, and $p_0$ is the value of Eqn. (\ref{an_pressure_final}) at $\psi=0$. One term $\cos\psi$ comes from Eqn. (\ref{an_pressure_final}), one from the surface element $2\pi R^2\cos\psi d\psi$, and one from the lever arm $R\cos\psi$.
    
    Introducing the non-dimentional TYORP torque $\tau_\omega=\frac{c}{\Phi R^3}T_\omega$ and performing integration in Eqn. (\ref{torque_dimentional}), we arrive at the following final expression for TYORP of the asteroid:
        \begin{equation}
        \tau_\omega= (1-A)p_0 \frac{8 \pi}{3}
        \label{torque_non-dimentional}
        \end{equation}
    
    Let us apply this formula to estimate TYORP of asteroid Ryugu. We assume the rotation period $7.63$ h, the albedo $A=0.045$ \citep{watanabe19}, the thermal emissivity $\epsilon=0.9$ \citep{hamm19}, the density of the subsurface material equal to the mean asteroid density $\rho=1190$ kg m$^{-3}$ \citep{watanabe19}, the thermal inertia $\Gamma=\sqrt{\kappa C\rho}=225$ J m$^{-2}$s$^{-0.5}$K$^{-1}$ \citep{shimaki2020}, and the heat capacity $\kappa=700$ J kg$^{-1}$ K$^{-1}$, which is typical for chondrite material \citep{szurgot11}. Substituting these values into Eqs. (\ref{L_cond}) and (\ref{theta}), we get the thermal parameter $\theta=1.34$ and the thermal wavelength $L_\mathrm{wave}=1.8$ cm. 
    
    For this thermal parameter, the maximal TYORP pressure is $p\approx0.0048k^2$, and the maximum is attained at the non-dimensional wavelength $l\approx4$ (see Fig. \ref{fig:tyorp}). It corresponds to the dimensional wavelength $L=L_\mathrm{wave}l\approx 7$ cm. The slope at this wavelength is $k \approx 0.5$ (see Figure \ref{fig:k}). It results into the maximal non-dimentional TYORP pressure $p_0\approx0.0012$.
    Substituting this value into Eqn. (\ref{torque_non-dimentional}), we obtain the non-dimensional TYORP produced by the surface roughness on Ryugu equal to $\tau\approx0.01$.
    
    \cite{kanamaru11} used the shape model of Ryugu revealed by \textit{Hayabusa2} to compute the normal YORP. The resulting acceleration was negative and lied in the range $-0.42 \cdot 10^{-6}$ deg day$^{-2}$ to $-6.3\cdot 10^{-6}$ deg day$^{-2}$. Such a large discrepancy between different shape models even at a very high resolution is not surprising. Extreme sensitivity of YORP to fine details of the shape models have been theoretically investigated by \cite{statler09}. Similar YORP differences for shape models of different (but very high) resolution have been also observed in simulations of asteroid Itokawa \citep{scheeres07}. Anyway, Ryugu's accelerations obtained by \cite{kanamaru11} correspond to the non-dimentional torque $\tau_\omega$ of the normal YORP in the range $-0.00014$ to $-0.0021$. The tangential YORP $\tau_\omega\approx 0.01$ derived from our model can surpass this value, causing a positive net acceleration despite the negative YORP produced by the asteroid shape asymmetry.
    
    \section{Discussion}
    \label{sec:discussion}

    Here, we constructed the theory of TYORP of a rough surface in application to the asteroid's regolith. But the same theory is applicable for the surface roughness of boulders, which can be described by the same equations albeit with different thermal parameters. Such surface roughness adds up to the torque created by the smooth boulders and can enhance TYORP. The largest TYORP is produced by the shape irregularities at the spatial scale of the order of $L_\mathrm{wave}$, which can correspond to small boulders as a whole as well as to surface roughness on large boulders. For a typical largely cracked protrusion, it can be impossible to tell whether it is an individual boulder or a bump on the surface of a larger boulder. This makes a smooth transition between TYORP produced by boulders and by surface roughness. At some degree of accuracy it can be even possible to express TYORP of gently sloped boulders in terms of this article by Fourier-decomposing the boulder shapes.
    
    For the typical asteroids' thermal inertias $\Gamma\sim 10-700$ J m$^{-2}$s$^{-0.5}$K$^{-1}$ \citep{hanus18}, densities $\rho\sim 1000-5000$ kg m$^{-3}$, rotation periods $P_\mathrm{rot}\sim 2-30$ h, and orbital semimajor axes $a\sim 1-3$ AU, we get the thermal parameters $\theta\sim 0.01-10$ and thermal wavelengths $L_\mathrm{wave}\sim 10^{-4}-10^{-1}$ m. The resulting thermal parameters are close to the values that produce the maximum TYORP (see Figure \ref{fig:tyorp}). The thermal wavelengths give us the scale at which the roughness produces the largest contribution to TYORP. Unfortunately, the information about roughness of asteroids at such a short scale is very limited. Moreover, it is questionable whether at sub-millimeter scale the asteroid material can be considered a continuous medium, thus violating the underlying assumptions of our model.
    
    If the rough surface elements have high tilts with respect to the horizon, they are shadowed more often than Eqn. (\ref{an_alpha}) assumes, which diminishes the TYORP produced by them. 
    Other intrinsic limitations of the proposed theory include the possible non-uniformity of the granular material at the $L_\mathrm{wave}$ scale, as well as the possible non-constancy of $\kappa$ as a function of temperature if a large fraction of energy is transferred by radiation (see e.g. \cite{ryan20}). These effects could be incorporated in the theory at a later stage.
    
    The most importantly, the suggested analytic theory needs to be verified by numeric simulations. But we spare this part for the subsequent paper.
%-------------------------------------------------------------------   
\section*{Acknowledgements}
\label{sec:acknowledgements}
This work had to be partially funded by the National Research Foundation of Ukraine, project N2020.02/0371 ``Metallic asteroids: search for parent bodies of iron meteorites, sources of extraterrestrial resources''. We are grateful to Ukrainians who are fighting to stop the war so that we can safely finish the revision of this article. We are also grateful to our reviewer Masanori Kanamaru for comments that helped to improve the article.

%-------------------------------------------------------------------

\end{document}